\documentclass[sigconf]{acmart}

\copyrightyear{2024}
\acmYear{2024}
\setcopyright{rightsretained}
\acmConference[ICSE '24]{2024 IEEE/ACM 46th International Conference on Software Engineering}{April 14--20, 2024}{Lisbon, Portugal}
\acmBooktitle{2024 IEEE/ACM 46th International Conference on Software Engineering (ICSE '24), April 14--20, 2024, Lisbon, Portugal}\acmDOI{10.1145/3597503.3623337}
\acmISBN{979-8-4007-0217-4/24/04}

\AtBeginDocument{%
  \providecommand\BibTeX{{%
    \normalfont B\kern-0.5em{\scshape i\kern-0.25em b}\kern-0.8em\TeX}}}

\usepackage{indentfirst}

\usepackage{xcolor,colortbl}

\usepackage[utf8]{inputenc}
\usepackage{algorithm}
\usepackage{algorithmic}

\usepackage{tabularx}

\usepackage{makecell}
\usepackage[export]{adjustbox}
\usepackage{xcolor}
\usepackage{float}
\usepackage{color}
\usepackage{tabularx}
\usepackage{numprint}
\usepackage{paralist}
\usepackage{balance}
\usepackage{amsmath}
\usepackage{subcaption}
\usepackage{tablefootnote}
\usepackage{soul}

\usepackage{hyperref}
\usepackage{graphicx,multirow}
\usepackage{newfloat,caption,subcaption,listings,mdframed,numprint}
\usepackage[font=small]{caption}
\usepackage{xspace}

\newcommand{\ICSERevision}[1]{\textcolor{black}{ #1}}

\usepackage{tikz}

\newcommand*{\Comb}[2]{{}^{#1}C_{#2}}%

\newlength\myindent
\definecolor{bluekeywords}{rgb}{0.13,0.13,1}
\definecolor{greencomments}{rgb}{0,0.55,0.2}
\definecolor{redstrings}{rgb}{0.9,0,0}

\newcolumntype{b}{X}
\newcolumntype{s}{>{\hsize=.5\hsize}X}

\newcommand{\approach}{{ITER}\xspace}

\definecolor{Gray}{gray}{0.92}
\definecolor{LightCyan}{rgb}{0.9,1,1}
\newcolumntype{a}{>{\columncolor{Gray}}c}
\newcolumntype{b}{>{\columncolor{LightCyan}}c}

\newcolumntype{g}{>{\columncolor{Gray}}l}

\usepackage{pifont}

\DeclareFloatingEnvironment[fileext=frm,placement={!ht},name=Listing]{listing}
\usepackage{tikz}

\title{ITER: Iterative Neural Repair for Multi-Location Patches}

\author{He Ye}
\email{hey@cs.cmu.edu}
\affiliation{
  \institution{Carnegie  Mellon University}
  \country{United States}
}

\author{Martin Monperrus}
\email{monperrus@kth.se}
\affiliation{
  \institution{KTH Royal Institute of Technology}
  \country{Sweden}
}

\begin{document}

\begin{abstract}
Automated program repair (APR) has achieved promising results, especially using neural networks.
Yet, the overwhelming majority of patches produced by APR tools are confined to one single location.
When looking at the patches produced with neural repair, most of them fail to compile, while a few uncompilable ones go in the right direction. 
In both cases, the fundamental problem is to ignore the potential of partial patches.
In this paper, we propose an iterative program repair paradigm called \approach founded on the concept of improving partial patches until they become plausible and correct. 
First, \approach iteratively improves partial single-location patches by fixing compilation errors and further refining the previously generated code.
Second, \approach iteratively improves partial patches to construct multi-location patches, with fault localization re-execution. 
 \approach is implemented for Java based on battle-proven deep neural networks and code representation.
 \approach  is evaluated on 476 bugs from 10 open-source projects in Defects4J 2.0.  \approach succeeds in repairing 15.5\% of them, including 9 uniquely repaired multi-location bugs.
\end{abstract}

\maketitle


\section{Introduction}
\label{sec:background}

Automated program repair (APR) \cite{TSE-repair-survey,Monperrus2015,goues2019automated} is a promising idea to incorporate different concepts from automated software engineering together in order to repair  software bugs automatically, eg. fault localization, code transformation and correctness verification. 
Test-suite-based program repair is one of the well-studied APR paradigms \cite{LeGoues2012GenProg,quixbugs-jss,sergey-test-equivalence-tosem18}, in which test-suites are considered as program correctness specifications.
Given a program and its test suite with at least
one failing test, APR approaches generate patches to make the test suite pass, such patches are known as \texttt{plausible patches}.
Finding plausible patches can be made with techniques involving search \ICSERevision{(e.g., GenProg \cite{genpro2009}, Elixir \cite{elixir} and Hercules \cite{hercules})}, semantic analysis \ICSERevision{(e.g., SemFix \cite{semfix}, DirectFix \cite{directfix} and Angelix \cite{Angelixicse16})}, and deep learning \ICSERevision{(e.g., SequenceR \cite{SEQUENCER}, SelfAPR \cite{selfAPR} and Modit \cite{modit-ase21})}.

\textbf{We note that all the aforementioned APR approaches repair bugs with a single repair attempt, a conceptual framework called in this paper the \textit{``one-step repair paradigm''}}.
\ICSERevision{
In this paper, a  "step" is defined as a complete repair attempt consisting of fault localization, patch generation, and patch validation, ultimately resulting in a decision to keep or discard such a patch. 
The term "one-step repair paradigm" signifies that the life cycle of the generated patch involves only one opportunity to be evaluated for keeping or discarding.
}
If there exists a patch to make all test pass, then that plausible patch is kept, otherwise, all intermediate patches are discarded. 
Different from one-step repair, multi-step repair that we propose in this paper consists of refining previous repair results.

\begin{figure}[t!]
\includegraphics[width=0.45\textwidth]{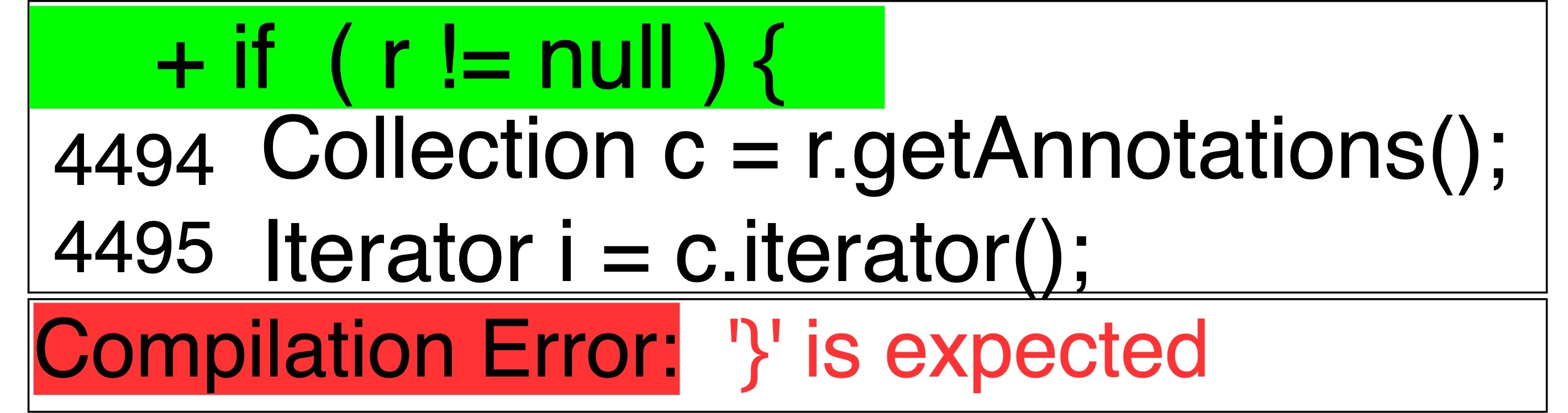} 

 \caption{A patch  for Chart-4 in Defects4J generated by AlphaRepair, Recoder, SelfAPR, RewardRepair, Cure, and CoCoNut. This compilation error patch could be further improved into a correct patch in a subsequent iteration. }
\label{fig:partial-ce-example}
\end{figure}

\textbf{Problem 1: One-step repair is 
inefficient for single-location bugs.}
The one-step repair paradigm ignores the potential of partial patches that could lead to correct patches with only a small adjustment.
To illustrate this point, \autoref{fig:partial-ce-example} shows a  patch that is commonly generated by six different repair techniques: AlphaRepair \cite{alpha-repair}, Recoder \cite{Recoder}, SelfAPR \cite{selfAPR}, RewardRepair \cite{RewardRepair-icse22}, Cure \cite{CURE-icse21} and CoCoNut \cite{CoCoNuT}. This patch applies a correct fix pattern on a correct buggy location. Unfortunately, all these techniques miss adding a '\}' to get a valid AST structure and complete the patch. 
This patch is partial: it goes in the right direction but it is not finished.
Yet, it is a good working solution that can be further improved into a correct patch. However, all existing APR approaches simply discard partial patches.


\textbf{Problem 2: One-step repair is fundamentally limited for repairing multi-location bugs.}
A multi-location patch is a patch that requires edits at multiple, non-contiguous locations in a program.
It is known that generating multi-location patches is especially challenging \cite{varfix}. There are only a few works target to repair multi-location bugs, with mixed success \cite{dear-icse22,varfix,hercules}. 
Repairing multi-location bugs needs to address two key challenges.
The first challenge is judging partial correctness in order to decide whether a work-in-progress patch should be discarded or further improved.
The second challenge is to correctly locate all buggy locations: most APR is founded on spectrum-based fault localization \cite{fl-tool} which only outputs single lines in isolation, and not tuples of lines that should be changed together.


To address the above two problems, we devise a novel program repair approach ITER, whose primary aim is to improve partial patches until they become correct.
\ICSERevision{
The novelty of ITER is to conduct an iterative patch refinement process,
coupled with iterative fault localization based on the partial patches obtained in previous repair steps. 
This patch refinement process leads to an updated test execution result, consequently improving the accuracy of fault localization.
To our knowledge, the iterative repair process has been demonstrated by only a few works such as
 GenProg \cite{genpro2009}, DeepFix \cite{deepfix} and an early work from Arcuri and Yao \cite{co-evolutionary-Arcuri-Yao-2008}.
ITER builds upon this iteration idea by introducing two unique components: 1) iterative patch refinement to keep and improve the intermediate patches while previous works choose to discard them and  2) iterative fault localization with the partial patches to obtain an updated list of suspicious locations.
}

\begin{figure}[t!]
\centering
\centering\includegraphics[width=0.4\textwidth]{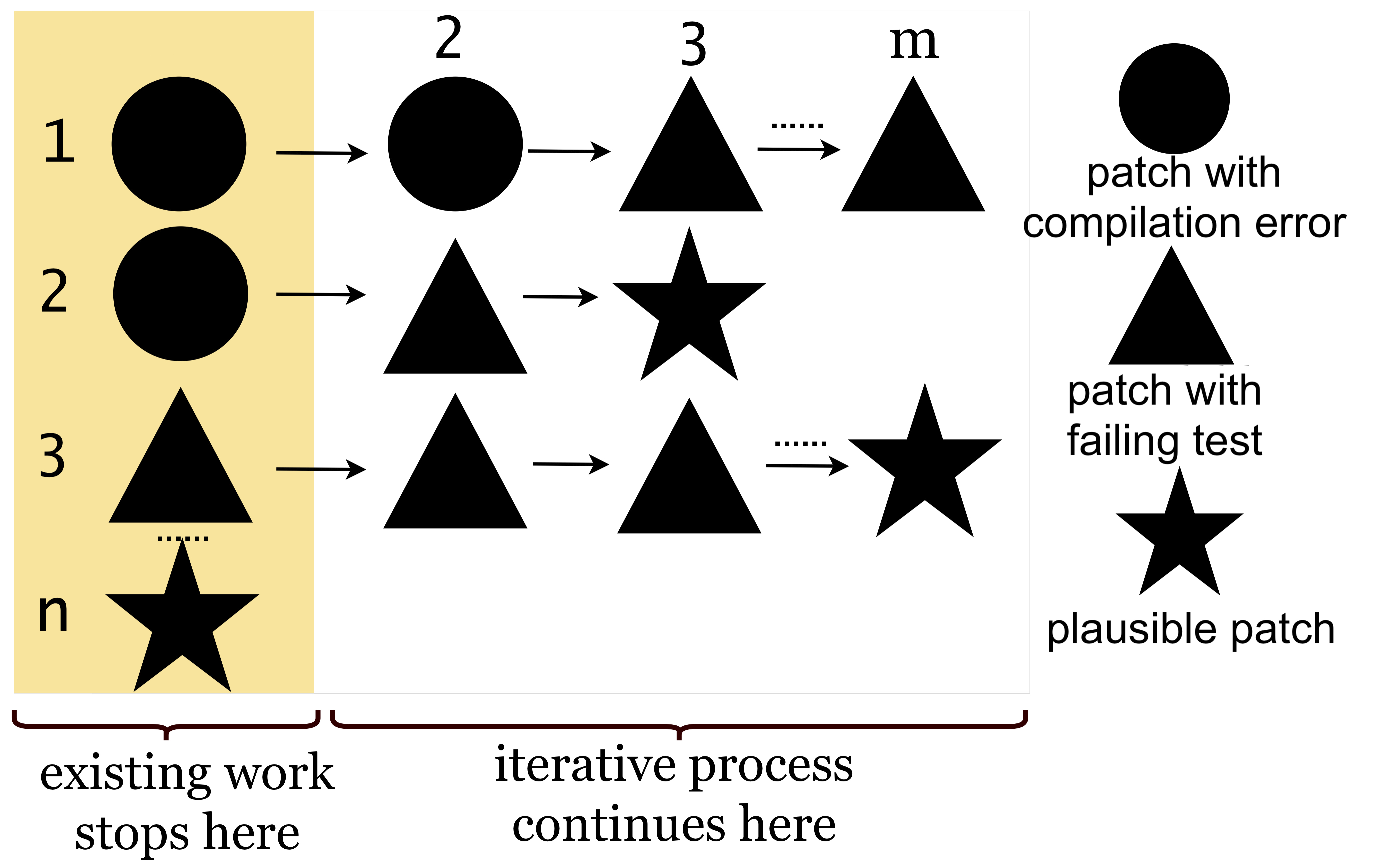} 
 \caption{Sketch of iterative repair: \approach chains compilation error repair and functional error repair \ICSERevision{with a single repair model}.}
\label{fig:motivating-sketch}
\end{figure}

\textbf{Solution 1:  Iteration over partial patches} 
\autoref{fig:motivating-sketch} shows a sketch of iterative repair. Existing work generates all candidate patches independently and without interaction, as highlighted in the yellow bar area. 
One-step repair would completely discard all non-plausible patches.
Departing away from one-step repair, \approach considers partial patches as the basis for further improvement. 
For example, in the second row of  
\ICSERevision{\autoref{fig:motivating-sketch}}
, \approach takes the first patch compilation error to further improve it, to turn it into a compilable patch with a test failure and then into a plausible patch that passes all tests. 
\approach chains compilation error repair and functional error repair in arbitrary order, as shown in the different rows of 
 \ICSERevision{\autoref{fig:motivating-sketch}}. 
\ICSERevision{The iteration over partial patches serves in both  training and  inference stages. The iterative training process not only enables the data augmentation with diverse training samples, but also enables the neural network to be trained with its own output, which subsequently becomes the new input in the next iteration.}

\ICSERevision{It is worth noting that prior work by Bhatia et al. \cite{chaining-repair} also explore  the concept of chaining compilation error repair and functional error repair. 
However, their approach utilizes two separate models, whereas ITER accomplishes this chaining using a single repair model. 
Additionally, ITER allows the flexibility of arbitrary orders for conducting compilation error repair and functional error repair, whereas the prior work strictly follows a sequential order of first addressing compilation errors and then fixing functional errors.
}

\textbf{Solution 2: Iteration over locations to be fixed}
\approach defines a way to iterate over fault localization results in a novel manner.
The core idea is to re-execute fault localization if a partial patch is found that reduces the number of failing tests. 
This results in an improved ranked list of suspicious buggy locations, updated per the behavioral changes from the partial patch. 
This is different from all existing repair approaches that only execute fault localization once. Our experiments give clear evidence that this is key to pushing the state-of-the-art of repairing multi-location bugs.

\begin{figure}[t!]
\centering
\includegraphics[width=0.5\textwidth]{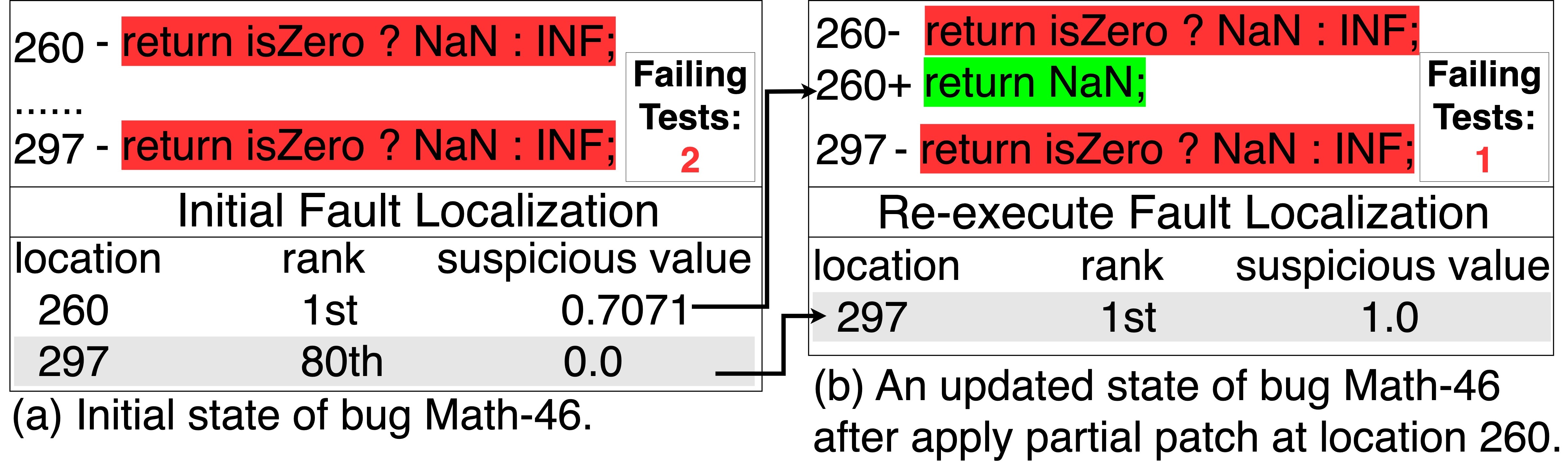} 
 \caption{Math-46: iterative execution of fault localization enables to focus on the next location to be repaired.}
\label{fig:partial-fl-example}
\end{figure}

Let us look at \autoref{fig:partial-fl-example} to give a better intuition of this 
\ICSERevision{iteration}
idea.
\autoref{fig:partial-fl-example} shows a bug from \texttt{Math-46} from Defects4J, which requires edits at two locations \texttt{line 260} and \texttt{line 297}.  A partial patch can be generated at \texttt{line 260} reducing the failing tests. Re-executing fault localization largely changes the suspicious lines and re-ranks the next buggy location \texttt{line 297} from the previous 80th place to the 1st place. This is explained as many suspicious locations are eliminated thanks to the partial patch being found, and the new ranked list only contains the most relevant suspicious locations.

We implement \approach on top of state-of-the-art neural program repair. 
This is very natural for two reasons.
First, our insight is that the same neural network can be good at fixing both compilation errors and functional errors, hence enabling the chained patch improvements of \autoref{fig:partial-ce-example}.
Second, one can use self-supervised training sample generation to train the network on a sequence of patches being improved from one to another and leverage the power of advanced input representations with error messages \cite{selfAPR}.

Our vision of iterative repair is multifaceted. 
At  training time,  \approach proposes an iterative loop by collecting the output of the patch generator to augment the training samples. This step largely increases the number of training samples: \numprint{2726077} new training samples, 5X more than the initial training dataset.
More importantly, \approach learns from its own mistakes, the failed repaired attempts in the previous optimization round.

At  inference time, \approach does what we call `iterative inference'', 
chaining fault localization, patch generation, and patch validation into one single 
repair loop. 
Patch validation executes tentative patches and provides the essential signals of failing tests and error diagnostics. 
Fault localization provides a ranked list of suspicious locations for the patch generator, based on the current partial patch. 
The patch generator produces candidate patches to repair either the initial bug or the current partial patch.
This happens in a loop until a plausible patch is found or a maximum budget is reached.

We evaluate \approach on 476 real-world bugs from 10 open-source projects collected in Defects4J \cite{defects4j} benchmark and compare \approach with  the recent relevant techniques.
\approach correctly repairs 74 of 476 bugs in the evaluation dataset.
Our experimental result shows \approach is 
able to improve partial patches into correct patches.
 \ICSERevision{
 As a result,
\approach repairs 9 unique multi-location bugs compared to related work Hercules \cite{hercules}, VarFix \cite{varfix} and  DEAR \cite{dear-icse22}, thanks to the effectiveness of iterative inference.
}

To sum up, we make the following contributions: 

\begin{itemize}
\item  We devise ``iterative repair'', a novel program repair paradigm for partial patch improvement. 
To our knowledge, this is the first repair architecture that chains repairs of compilation errors and test errors \ICSERevision{with one single repair model}, refining partial patches until they become plausible.

\item We implement \approach, realizing the vision of iterative repair for Java, based on 
\ICSERevision{state-of-the-art}
deep learning models and input/output representations.

\item We perform an original series of experiments and show that  \approach repairs 74 bugs from 10 open-source projects in Defects4J. Per our goal, we demonstrate that  \approach is able to repair multi-location bugs, better than the state-of-the-art, with 
\ICSERevision{9}
patches that have never been synthesized.

\item We make all our code and data publicly available at
\color{blue}{\url{https://github.com/ASSERT-KTH/ITER}} \color{black}{ and we have set up an online demonstration at}
 \color{blue}{\url{https://www.iterativerepair.tech} }.

\end{itemize}

\section{Approach}

\begin{figure*}[t]
\includegraphics[width=0.85\textwidth]{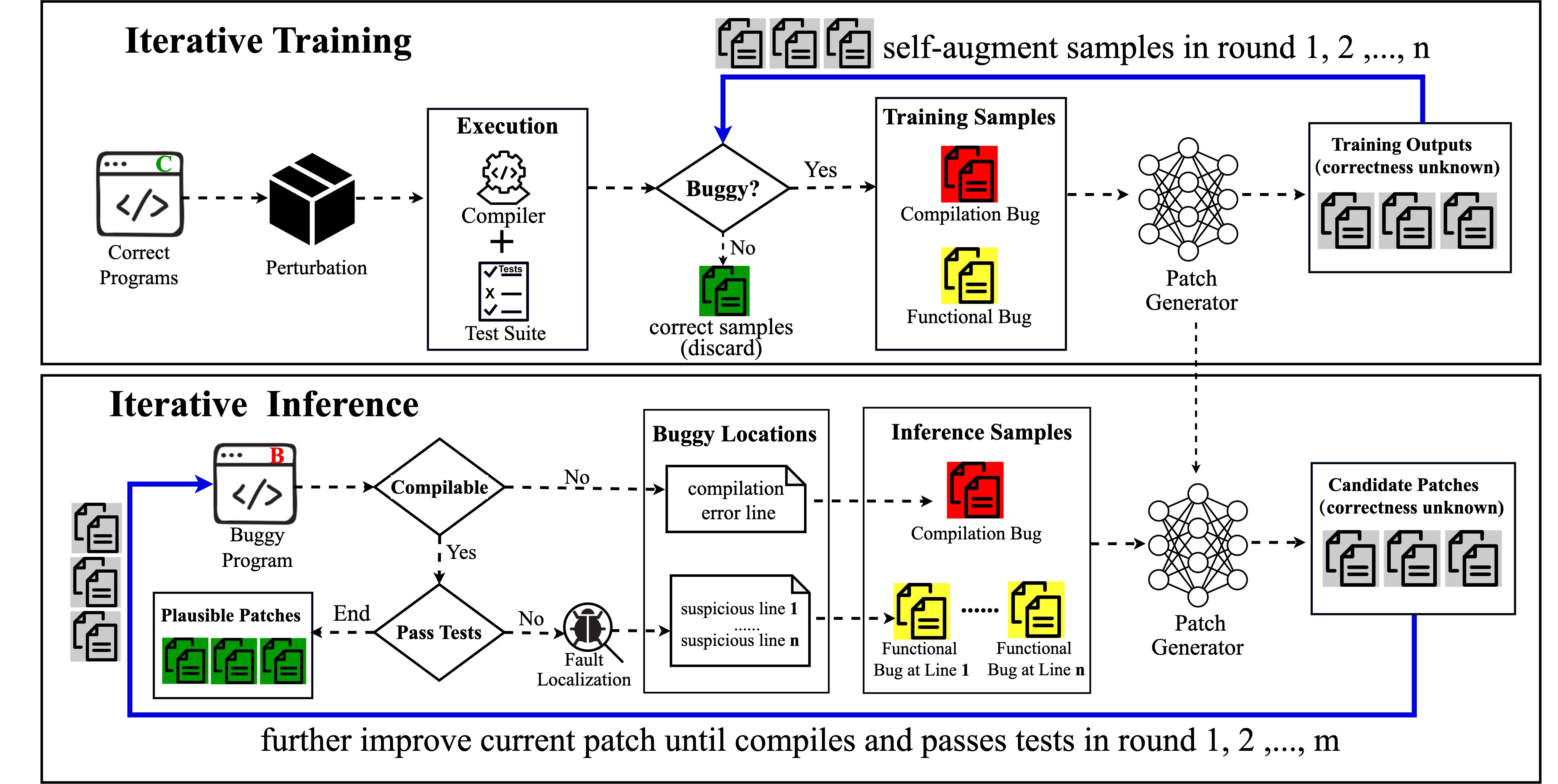} 
 \caption{Overview of \approach: Iterative Training generates valuable training data in a self-supervised manner, Iterative Inference chains compilation error repair and functional error repair to improve partial patches until they become plausible.}
\label{fig:overview}
\end{figure*}

\autoref{fig:overview} gives an overview of \approach. 
\approach has two main novel concepts: iterative training and iterative inference.
The upper part illustrates the iterative training  and the lower part shows the iterative inference.  As highlighted in the arrows in blue, 
 both the training and inference consider the output from the patch generator as the basis of the new bugs and further improve them to seek an iteration loop.

At training time, the patch generator at a given iteration $1,..,n$ is used to generate self-augment samples in an iterative loop.  The number of iterations in the loop is configurable by end users and this iteration ends when the max iteration configuration is achieved. 
There are two insights behind iterative training. 
First, this self-augmentation increases the number of training samples beyond the initial training dataset.  
Second, such an iterative loop enables the patch generator to learn to repair its own mistakes (i.e., its output). 

At inference time, the iterative vision of \approach is as follows: all candidate patches generated by the patch generator that are not yet plausible are further improved  until they compile and pass the test specification.  
\approach conducts the following three steps: 1) fault localization, 2) patch generation, and 3) patch validation. 
For each iteration, \approach  keeps track of the state of the generated patch with three pieces of information:
1) the new diagnostic (compiler error or failed test error message). 
2) the new buggy locations that are identified for uncompilable patches (from a compiler).
3) the new suspicious locations that are identified with fault localization (from fault localization).

\approach is novel in the following:
First, \approach conducts the iteration process in both training and inference processes. 
\approach considers improving two types of patches with a) compilation errors or b) function errors until they compile and pass test cases.
This is in contrast to all the existing step repairs in the literature \cite{alpha-repair,xia-llm-icse23,Recoder,varfix,jiang2023knod} that only conduct a one-step repair.
Second, \approach conducts the iteration loop by integrating three repair components into one iterative process: fault localization, patch generation, and patch validation. 
This is different from existing  learning-based  repair techniques that consider fault localization and patch validation as separate pre-process and post-process tasks from patch generation.
This is different from all the prior work that only executes fault localization once. 

\subsection{Iterative Training}
\label{sec-iterative-trianing}
\approach is built on the concept of self-supervised learning which consists of automatically producing labeled data and bug diagnostics for unlabeled programs. 
In the context of repair, it has been proven useful for grounding training on the execution of the buggy and patched programs \cite{selfAPR,allamanis2021self-buglab,Yasunaga20DrRepair}. 

In \approach, self-supervised learning is used to provide two kinds of training samples: compilation bugs and functional bugs as well as the corresponding fixes. This is the key to iteratively improving the different patches until they compile and pass all tests, potentially alternating between compilation errors and functional errors.
Second, self-supervised learning with execution enables \approach to include diagnostics of  buggy programs to be repaired, which is important guiding information to drive the iterative repair process \cite{selfAPR}.

\textbf{Input and Execution} For training, \approach takes an input of correct programs and perturbs them into buggy programs using a perturbation model \cite{selfAPR,allamanis2021self-buglab,Yasunaga20DrRepair}.
To determine the bugginess of the perturbation-based samples, per the prior work \cite{selfAPR},  \approach executes perturbation-based samples  with the compiler and the test suite,  and \approach only keeps those samples that cause compilable errors or test failures. 
All buggy  samples are fed to a neural patch generator whose goal is to learn to transform a buggy program ($b$) into the fix program ($f$).

\textbf{Iteration} At a given point in time, \approach's neural network outputs different variants of the bug $\mathcal{V}$ (i.e., candidate patches) for a given buggy sample based on a configurable beam search width $k$.
The usage of those training outputs is two-fold. 
First, they are used to compute a loss value and to optimize the neural network with back-propagation accordingly. 
Second, they are used as new training samples (as shown in the blue arrow), as a data augmentation technique. 
These self-augment training samples generated from the neural networks are able to increase the number of training samples beyond the perturbation model that is used. 
Importantly, the neural network is trained to repair its own mistakes and this is intuitively useful for \approach to repair real-world bugs at inference time.

\textbf{Iterative Training Sample Selection}
Algorithm \autoref{alg:training} shows the generation and selection of new training samples.
\approach takes as input the initial training samples  $\mathcal{I}$ consisting of pairs of buggy and ground truth code \texttt{($b_{i}$,$f_{i}$)} and a neural patch generator $\mathcal{G}$.
\approach is configured be a beam size width as \texttt{k} and a max iteration \texttt{MAX}.
As shown in line 7, for each buggy sample \texttt{$b_{i}$},  the patch generator outputs a set of variants $\mathcal{V}$  of \texttt{$v_{i}$} based on beam size width \texttt{k}. Not all the $v_{i}$ are considered valid augmented training samples to be added to $\mathcal{S}$. We discard those that identical to the buggy program \texttt{$b_{i}$} and those that are identical to the ground truth fix \texttt{$f_{i}$} (in line 10 and 11). All the iterative training samples $\mathcal{S}$ are combined with initial training input $\mathcal{I}$ and they form the new training dataset for the next iteration.

\subsection{Iterative Inference}
The core idea of \approach is to iteratively improve the current patches until they compile and pass tests. \approach considers a partial patch as the basis for the next improvement.
\approach integrates fault localization, patch generation and patch validation in each iteration loop rather than treating them as three separate tasks as related work \cite{alpha-repair,Recoder,RewardRepair-icse22,selfAPR,CURE-icse21,CoCoNuT,xia-llm-icse23}. 
Such integration brings two benefits. 
First, patch validation selects partial patches that are able to  improve the effectiveness of fault localization (as demonstrated in \autoref{fig:partial-fl-example}). Second and accordingly, a better fault localization further improves the  effectiveness of \approach in generating correct patches.

\textbf{Input and Execution}
At inference time, \approach takes as input a buggy program and executes it against the accompanying compiler and test suite to determine whether it is a compilation bug or a functional bug, and to collect the diagnostic. 
If the bug contains compilation errors, \approach takes the explicit compilation error line and diagnostic based on the compiler. 
If the bug contains functional errors, \approach executes  fault localization  to produce a ranked list of suspicious locations, each of which is considered a separate bug to be repaired.

\begin{algorithm}[t!]
\tiny
  \caption{Iterative Training}
  \begin{algorithmic}[1]
   \STATE \textbf{Input} initial training samples $\mathcal{I}$: ($b_{1}$,$f_{1}$), ($b_{2}$,$f_{2}$),...,($b_{n}$,$f_{n}$), patch generator: $\mathcal{G}$
    \STATE \textbf{Config:} beam search width: k, max iteration:  MAX
   \STATE  \textbf{Output:} iterative training samples $\mathcal{S}$ $\gets  \emptyset $

    \STATE {$\mathcal{S}$  $\gets$ $\mathcal{I} $}
    \STATE {count $\gets$  0 }
    \WHILE{count < MAX}
     \FOR{($b_{i}, f_{i}$) in $\mathcal{S}$} 
      \STATE{ $\mathcal{V}$ : $v_{0}$ ... $v_{k}$ 
      $\gets$ $\mathcal{G}$ ($b_{i}, k$)}
       \FOR{$v_{i}$ in $\mathcal{V}$} 
        \IF{$v_{i}$ $!=$ $b_{i}$ $\AND$ $v_{i}$ $!=$ $f_{i}$ }

        \STATE{  $\mathcal{S}$ $\gets$ $\mathcal{S}$ $\cup$  ($v_{i}$, $f_{i}$) }
        \ENDIF
       \ENDFOR
     \ENDFOR
          \STATE{ count++}
     \ENDWHILE

  \end{algorithmic}
  \label{alg:training}
\end{algorithm}

\textbf{Iteration}
\approach outputs different candidate patches for a given bug based on the configurable beam size $k$.
All the candidate patches are tested against the compiler and test suite.  If those patches  are buggy (uncompilable or failing tests), they are taken again as input in order to be further improved, this is the core idea of iterative inference. 
After one or more iterations, if one patch is plausible, i.e., it passes all test cases, it goes to the plausible patch pool and it will not be further improved. 

The iteration is signaled by the number of failing test cases. 
There are two options to be considered during iteration.
Option 1: when a partial patch is found to reduce the number of failing tests, then this patch is kept and the fault localization is re-executed based on this partial patch.
Option 2: when no partial patch reduces the number of failing tests, \approach continuously improves the buggy location in its last iteration. 




\textbf{Iterative Algorithm}
Algorithm \autoref{alg:inference} explains the iterative inference process as pseudo-code. The key recursive repair function is called in \texttt{line 29} by providing the newly generated patch \texttt{p} based on the previously patched program \texttt{b'}.
The patch \texttt{b'} is validated against compiler $\mathcal{C}$ and test suite $\mathcal{T}$ to obtain the  current fail tests $\mathcal{N}_{b'}$, note that $\mathcal{N}_{b'}$ is only obtained when \texttt{b'} is compilable (line 7).    
\approach takes the failing test number as the signal to identify partial patches to be kept and re-execute the fault localization $\mathcal{F}$. As shown in line \texttt{line 22 } and \texttt{line 23}, fault localization $\mathcal{F}$ is re-executed when $\mathcal{N}_{b'}$  is fewer than initial failing tests $\mathcal{N}_{b}$.
\approach is ended under two conditions. First, \approach is ended when a plausible patch \texttt{p} is found (\texttt{line 17}), which forms the pair of data $<p,\mathcal{L}>$.
Second,  \approach is ended when  iteration $i$ achieves a maximum configuration (\texttt{line 20}).

In theory, the maximum number of total candidate patches $P$ on one location $\mathcal{L}$ depends on the beam search width $k$ (i.e. the number of candidate patches generated per bug) and the number of iteration $i$. We compute the candidate patch number $P$  in \autoref{equ-patch-number}. 
\begin{equation}\label{equ-patch-number}
\footnotesize
    P  =  k^{i}
\end{equation}

\begin{algorithm}[t!]
\tiny
  \caption{Iterative Inference }
  \begin{algorithmic}[1]
   \STATE \textbf{Input:} initial buggy program $b$,  fault localization $\mathcal{F}$, patch generator $\mathcal{G}$, compiler $\mathcal{C}$, test suite $\mathcal{T}$,  initial failing tests number  $\mathcal{N}_{b}$
   \STATE \textbf{Config:} beam search width $k$, max iteration MAX 
   \STATE \textbf{Output:}   plausible patches $plausible$
    \STATE \textbf{Init:}  $plausible \gets \emptyset$,

    \STATE
    \STATE \textbf{MAIN()}
     \begin{ALC@g}
        \STATE $p \gets \emptyset$  //initial patch
         \STATE $i$ $ \gets$ 0  //iteration
       \STATE $\mathcal{L} \gets \emptyset$ //buggy location
    \STATE \textbf{ITER} (b, p,  k, i, $\mathcal{F}$, $\mathcal{G}$, $\mathcal{C}$, $\mathcal{T}$,  $\mathcal{L}$, $\mathcal{N}_{b}$)
    \end{ALC@g}

   \STATE
    
    \STATE \textbf{ITER} (b, p,  k, i, $\mathcal{F}$, $\mathcal{G}$, $\mathcal{C}$, $\mathcal{T}$, $\mathcal{L}$, $\mathcal{N}_{b}$)
        \begin{ALC@g}
       \STATE b' $\gets$ apply(b, p)           
    //apply patch

      \STATE  $\mathcal{N}_{b'}$  $\gets$ validate(b', $\mathcal{C}$, $\mathcal{T}$)

      \IF{$\mathcal{N}_{b'}$ == 0}
      \STATE $plausible$ $\gets$ <p,$\mathcal{L}$>
      \STATE break   //break repair iteration when plausible patch is found

      \ENDIF

      \IF{i > MAX}
        \STATE break  {  //end repair iteration}
      \ENDIF
       
       \IF{ $\mathcal{L}  \in \emptyset$ or $\mathcal{N}_{b'}$ < $\mathcal{N}_{b}$  }
       \STATE   $\mathcal{L}$  $ \gets$  $\mathcal{F}$ ( b', $\mathcal{C}$, $\mathcal{T}$)   // fault localization iteration
       \STATE $\mathcal{N}_{b}$  $\gets$ $\mathcal{N}_{b'}$ 
       \ENDIF

       \FOR{loc in $\mathcal{L}$}
        \STATE $\mathcal{P}$  $\gets$   $\mathcal{G}$ (b', loc, k)
         \FOR{p in $\mathcal{P}$}
         \STATE \textbf{ITER} (b', p, k, i++, $\mathcal{F}$, $\mathcal{G}$, $\mathcal{C}$, $\mathcal{T}$, loc, $\mathcal{N}_{b}$)   // repair iteration        
          \ENDFOR
       \ENDFOR
       \end{ALC@g}

  \end{algorithmic}
  \label{alg:inference}
\end{algorithm}



\section{Experimental Methodology}
\subsection{Research Questions}

\begin{itemize}
\item \textbf{RQ1 (Effectiveness)}: To what extent does \approach perform compared to related work?
\item \textbf{RQ2 (Multi-location Bugs)}: To what extent does \approach repair multi-location bugs?
\item \textbf{RQ3 (Ablation Study)}: To what extent does each component in \approach contribute to the final effectiveness?
\item \textbf{RQ4 (Time Cost)}: What's the time cost of \approach in generating plausible patches? 
\end{itemize}

\begin{table}[t!]

\renewcommand{\arraystretch}{1.0}
\footnotesize
\caption{Iterative Training Dataset. In each iteration, we execute \approach to generate two patches per bug and keep the valid ones for the next iteration.}

\begin{tabular}{ccc}

\hline

Training Iteration & Self-augment Samples & Total \\

\hline
Initial  & - & \numprint{651787} \\

Iteration\_1 &+\numprint{541192} &   \numprint{1192979}  \\
Iteration\_2 &+\numprint{879012} &   \numprint{2071991}  \\
Iteration\_3 &+\numprint{1305873} &  \numprint{3377864}    \\
\hline
Total Training Samples for \approach & +\numprint{2726077} &  \numprint{3377864}   \\

\hline
\end{tabular}


\label{table-training-dataset}
\end{table}

\subsection{Implementation}
\textbf{Training Dataset} We implement \approach on the data from Ye et al. \cite{selfAPR} comprising \numprint{651787} patches, including  \numprint{281762} compilation error samples and  \numprint{370025} functional error sample. Recall that it is key for us to have both kinds of errors  to enable iterative repair. \approach follows the same code representation as \cite{selfAPR} by considering the context code surrounded by buggy code, additional meta-information (e.g., variables grouped by type), and diagnostics of the bug.  \approach follows the prior with to split compilation error diagnostic and functional error diagnostic with special tokens $[CE]$ and $[FE]$.
We configure \approach to take a maximum of $384$ input tokens as context code and generate a patch with a maximum of $76$ tokens with a batch size of $32$ and a vocabulary size of \numprint{32128}.

\approach is configured to be trained with three iterations, i.e., \texttt{$MAX=3$}, and the beam search width \texttt{$k=2$}. This means that \approach generates two candidate patches for each single given buggy location and this continues for three iterations.
\autoref{table-training-dataset} gives the self-augment training dataset generated by \approach during an iterative loop. In total, \approach is trained on \numprint{3377864} training samples including \numprint{2726077}  self-augment samples generated with our novel iterative loop.
This means that \approach's iterative training results in 5X more training samples than simple self-supervised data generation per \cite{selfAPR} (\numprint{3377864}  vs. \numprint{651787}). 
We observe that the number of self-augment training samples is not exactly twice than that of the training samples in the last iteration. This is because \approach discards those generated patches that are identical to the buggy and ground truth fix (see \ICSERevision{Algorithm} \autoref{alg:training}).

\begin{table*}
\renewcommand{\arraystretch}{1.2}
\scriptsize
\caption{Comparison with State-of-the-art. In the cells, x/y : x denotes the number of correct patches,
and y denotes the number of plausible patches that pass all human-written test cases. A ‘-’ indicates that the number has not been reported in the corresponding article.}

\begin{tabular}
{p{0.12\linewidth}p{0.03\linewidth} p{0.01\linewidth} 
p{0.07\linewidth}p{0.07\linewidth}ccc
aaaa}

\hline
  \\
Projects &Bugs &   & SimFix & TBar   & Recoder & AlphaRepair & SelfAPR  &  \multicolumn{4}{c}{\approach} \\


& &     & && &&&    &1 &2 & 3  \\

\hline

Chart & 26 && 4/8 & 9/14   &  8/14 &  6/-    &   6/10  & &  6/10 & 8/11  & 9/13 \\
Cli &39 &&  0/4  &  1/7  &  3/3     & 5/5 &2/7  &  &  3/7 &  5/11  &  6/13  \\
Closure & 133 &&  6/8 & 8/12 &13/27   & 12/ & 12/20 &   &  11/18 & 13/21 & 15/22\\
Codec & 18 &&   0/2  & 2/6   & 2/2     &  6/7 &  3/6 &&  1/3 & 1/4 & 2/6 \\
Compress & 47 & & 0/6 & 1/2 & 1/3  & 1/3 & 1/2 &&  0/2 & 1/4 & 2/6 \\
Csv &16 && 0/2 & 1/5 & 4/4 & 1/2 & 0/1 & & 1/2 & 2/5 & 2/5 \\
JacksonCore & 26& & 0/0 & 0/6 & 0/4 & 3/3 & 1/3 && 1/3 & 2/3 & 2/3 \\
Lang &  65 && 9/13 & 5/14 & 9/15 &11/-& 6/12 & &  5/9  & 9/10  & 9/11 \\
Math &105 && 14/26 & 18/36 &15/30 &  16/- & 13/24  & &  12/23 & 21/32 & 25/36 \\
Time & 26 && 1/1 &1/3 & 2/2 &  3/-  & 1/3 &  & 1/2 & 2/4 & 2/4 \\
\hline
Total & 476 & & 34/70 & 46/105  & 57/104 &  64/- & 44/91  && 41/79 & 64/105 & \textbf{74/119}\\

\hline

\end{tabular}

\label{sota-comparison}
\end{table*}

\textbf{Inference}
\approach integrates GZoltar \cite{GZoltar} as fault localization to provide suspicious buggy locations to be repaired. 
\ICSERevision{
There are two reasons for choosing GZoltar as the fault localization tool. 
First, GZoltar is widely used in program repair related works, such as DEAR \cite{dear-icse22}, Recoder \cite{Recoder}, SelfAPR \cite{selfAPR}, SimFix 
\cite{Simfix:2018}, and TBar \cite{tbar}. Considering the same implementation and same suspiciousness formula Ochiai (default formula used  in GZoltar) enables us to conduct a fair comparison with the related works.
Second, the effectiveness of GZoltar in detecting real-world bugs has been demonstrated in previous comprehensive evaluations with different fault localization techniques  \cite{gzolta-justification-icse17}.
}
\approach is implemented with  GZoltar 1.7.2 \footnote{https://github.com/GZoltar/gzoltar/releases/tag/v1.7.2. Released on May 9, 2019}.  Specifically, we configure the fault localization with \texttt{ochiai} formula  and  \texttt{entropy} metric. The detailed setting can be found in our online repository.

\ICSERevision{
At inference time, 
ITER  at most processes the top-50 suspicious locations identified by GZoltar. This is a smaller number compared to  related work, such as Hercules \cite{hercules}, which processes  top-200 suspicious locations. However, considering the iterative nature of the repair process, ITER indeed generates more than 200 tentative patches at the end.
}
\ICSERevision{
\approach  configures the repair max iteration \texttt{$MAX=3$} and the beam search width \texttt{$k=10$}.
Consequently, \approach generates 10 patches in each iteration, and this results in at most $10^3=1000$  patches according to \autoref{equ-patch-number}.
Performing three iterations (\texttt{$MAX=3$}) is a reasonable number based on our experimental results (we demonstrate it in RQ1).  Beyond this iteration threshold, there will be a significantly increased computational overhead. 
While a common choice for the beam search width ranges from 50 to 100 (SequenceR \cite{SEQUENCER} and Recoder \cite{Recoder} consider a beam width of 50 and 100, respectively), ITER deliberately chooses a narrower beam search width of \texttt{$k=10$} to accommodate the iterative process. 
This sacrifice allows ITER to focus on the iteration process and enables early stopping when a partial or plausible patch is found, optimizing efficiency and reducing unnecessary computational overhead.
}

\subsection{Patch Assessment}
\label{sec:patch-assessment}

The patch Assessment follows the existing related work  \cite{SEQUENCER,CoCoNuT,CURE-icse21,Recoder,alpha-repair}. 
All the plausible patches are  manually analyzed based on the ground truth developer's patch. 
To sum up, a patch is deemed correct if 1) it is identical to the developer-provided patch, 2) it is identical to correct patches generated by existing techniques where the patches have been publicly reviewed by the community in an open-source repository, 3) it is semantically equivalent to the developer-provided patch.

\subsection{Methodology for RQ1}
In RQ1, we compare 
\approach against the state-of-the-art program repair baselines on 10 open-source projects from widely used benchmark Defects4J 2.0 \cite{defects4j}.
Since \approach is based on the core concept of iterative fault localization, we compare against techniques that also rely on fault localization, spanning over both learning-based and template-based APR. 
For learning-based APR, we choose three recently published approaches evaluated on Java bug datasets with fault localization:
Recoder \cite{Recoder},  SelfAPR \cite{selfAPR}, and AlphaRepair \cite{alpha-repair}. 
For template-based APR, we choose two best-performing techniques with fault localization according to Zhu et al. \cite{Recoder} evaluation:  TBar \cite{tbar} and SimFix \cite{Simfix:2018}.
Note that all papers \cite{CURE-icse21,CoCoNuT,RewardRepair-icse22,xia-llm-icse23,jiang2023knod,SEQUENCER} assuming perfect fault localization are naturally excluded from this comparison.  We report the quantitative results from the corresponding papers and repositories \cite{Liu2020Efficiency,Recoder,alpha-repair}.
We measure the number of plausible patches and correct patches according to the criteria discussed in Section
\ref{sec:patch-assessment}.



\subsection{Methodology for RQ2}

One of the main contributions of \approach is to integrate fault localization into an iteration loop and to re-execute  fault location based on partial patches.
This is the essential element to further improve partial patches with a better ranked list of unrepaired locations.
Therefore, we evaluate  \approach in repairing multi-location bugs.
For RQ2, we take a subset from Defects4J  \ICSERevision{2.0} \cite{defects4j} by considering the following criteria: 1) multi-class bug: the ground truth patch by the developers spread more than one buggy class. 2) multi-edit bug: the ground truth patch contains more than one  non-contiguous edit location in one buggy class. 3) the ground truth patch contains less than five edit locations in total. 
This gives us \ICSERevision{35}  multi-location bugs for evaluation.

We compare \approach against the state-of-the-art of multi-location repair approaches: Hercules \cite{hercules}, VarFix \cite{varfix} and \ICSERevision{DEAR \cite{dear-icse22}}.
Hercules repairs multi-location bugs by searching for siblings (i.e., similar-looking code).
VarFix repairs multi-location bugs by  merging
all potential edits into a meta-program, where they are guarded
by if-conditions that are controlled at runtime.
\ICSERevision{
To compare with DEAR \cite{dear-icse22}, we re-implement it because of the unavailability of the repaired bugs and patches in their open repository. 
Following the description provided in the paper, we re-implement DEAR's fault localization technique which consists of two steps: pair-wise hunk detection and multiple-statement expansion.
In the first step, a hunk detection model is trained to learn whether two statements from different hunks should be fixed together.
In the second step, DEAR trains a multiple-statement expansion model to determine whether the candidate locations determined in the first step should be expanded to their surrounding locations.
In our re-implementation, both the hunk detection and statement expansion models are trained based on DistilBERT \cite{sanh2020distilbert} \footnote{provided by Hugging Face: https://huggingface.co/distilbert-base-uncased}.
The training data for hunk detection learning consists of training pairs that combined from the top-50 ranked statements from GZoltar based on 430 Defects4J bugs.
On the other hand, the training data for statement expansion model consists of training points directly from those top-50 ranked statements.
In both training dataset setups, the human developer patch locations provided by Defects4J are considered as the ground truth of buggy.
The patch generator is ITER's one, which enables us to well isolate the effect of fault localization.
The  detail of re-implementation setup is given in our open-source repository.
}

\subsection{Methodology for RQ3}
In RQ3, we conduct an ablation study to evaluate the contribution of self-supervised training samples that are added at each training iteration.
Specifically, we measure performance with:  \approach with no additional sample (denoted as $\approach_{0}$), augmented with training samples in iteration 1 ($\approach_{1}$), augmented with training samples in iteration 2 ($\approach_{2}$).
We evaluate performance on Defects4J (v1.2) per the number of bugs that are generated with plausible patches.

\subsection{Methodology for RQ4}
In RQ4, we analyze the time cost spent in generating plausible patches at inference time. This time costs sum up the costs of calling the neural network, running the compiler and tests to identify partial patches, and re-running fault-localization.
Specifically, we analyze the time cost depending on the number of inference iterations and report the minimum, maximum, and median time cost of plausible patches generated by \approach.

\section{Experimental Results}
\subsection{Answers to RQ1: Effectiveness}

In RQ1, we compare the effectiveness of \approach with five state-of-the-art program repair techniques over 476 bugs from 10 open-source projects from the Defects4J v2.0 benchmark.
\autoref{sota-comparison} give the evaluation results. The first column gives the project name and the second column gives the number of bugs in this project for evaluation. 
From the third to the last columns, we report the evaluation results of correct patches and plausible patches separated by a slash (correct/plausible).
Specifically for \approach, we evaluate the repair results of  the \approach over all three iterations of iterative inference, which is indicated by the iteration number.

\approach successfully generates
correct patches for 74 bugs and  plausible patches for 119 bugs,  which outperforms all previous techniques over both template-based and learning-based APR techniques. \approach improves the next best baseline AlphaRepair \cite{alpha-repair} by 15.6\% (74 vs. 64) in terms of correct patch generation.  \approach improves the next best baseline TBar \cite{tbar} by 13.3\% (119 vs. 105) in terms of plausible patch generation.


\textbf{Improvement over iteration.}
The last three columns clearly show that \approach's effectiveness comes from our backbone concept of  iterative repair. \approach boosts the correct patch number from 41 in \texttt{iteration 1} to 74 in \texttt{iteration 3}, which is an 85.4\% improvement compared to only using the patch generator once. We now give two examples to illustrate how \approach generates correct patches based on partial patches that contain either a compiler error or a  functional error.

\begin{figure}[ht!]
\includegraphics[width=0.45\textwidth]{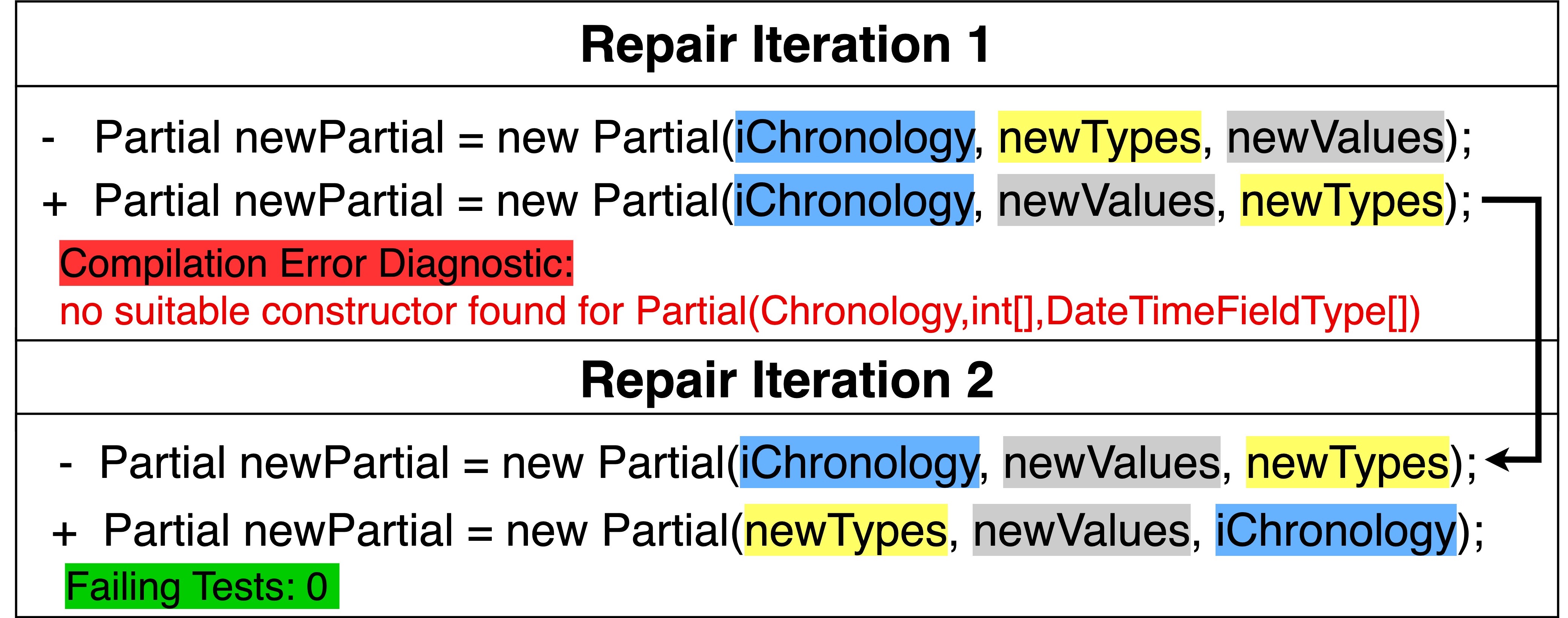} 
 \caption{\approach generates a correct patch for Time-4 based on improving a partial patch with a compiler error.}
\label{fig:ce2correct-example}
\end{figure}

\emph{A correct patch derived from a temporary uncompilable.} \autoref{fig:ce2correct-example} gives an example where \approach generates a correct patch for bug \texttt{Time-4} with two iterations, where the patch is identical to the ground truth fix provided by the developer. 
This patch requires a swap among three parameters in a method call. 
In the first iteration, \approach swaps the positions of only two parameters \texttt{newTypes} and \texttt{newValues}, and this leads to a compiler error: type checking fails because of the absence of the appropriate constructor. 
In the second iteration, \approach produces a patch based on the previous compiler error by swapping another set of parameters \texttt{iChronology} and \texttt{newTypes}. The second repair iteration leads to a correct patch. 
This patch is only found in the second iteration and consists of an iterative chain of compilation error repair and functional error repair.



\textbf{Analysis of patch evolution over iterations.}
We also analyze the evolutionary path of plausible patches generated by \approach.
Recall that for each bug, there are multiple candidate patches that could be generated by compounding beam and iterative improvement. \approach generates a total \texttt{1432} different plausible patches for 119 bugs, approximately 12 plausible patches per bug.  
\autoref{table-patch-evolution} visualize the evolutionary path for 1432 plausible patches.  In the table, the \texttt{CE} and \texttt{FE} respectively indicate compilation error and functional error, i.e. a test failure.

The first column gives the number of iterations and the total number of plausible patches found at this iteration.
The second column gives the evolution path in each iteration. For example, \texttt{CE->plausible} means that the first partial patch is with a compilation error and the improved patch in the second iteration is plausible.

Overall, we make the following observations.
First, each iteration indeed produces more plausible patches: $231 < 390 < 811$. Because of the compounding effect of beam and fault localization, the trend is clearly superlinear.  
Second, both temporary compilation errors (CE) and functional errors (FE) contribute to the final effectiveness of \approach. There are four patch evolutions (P2, P4, P5 and P6) that involve partial patches with compilation errors and yield 36.2\%  (7.2\% + 4.3\% + 1.3\% + 23.4\%) of the final plausible patches.

\begin{table}[t!]

\renewcommand{\arraystretch}{1.0}
\footnotesize
\caption{Evolution analysis of 1432 plausible patches. CE and FE respectively indicate compilation error and functional error.}
\begin{tabular}{cgcccccc}

\hline

No. Iteration&\multicolumn{1}{c}{ Patch Evolution} & Patches &  Percentage\\
\hline
1 (231)&P1: plausible & 231 & 16.2\% \\

\hline

\multirow{2}{*}{2 (390)}&P2: CE->plausible  & 103 & 7.2\%  \\

&P3: FE->plausible & 287 & 20.0\% \\
\hline

\hline

\multirow{4}{*}{3 (811)}&P4: CE->CE->plausible & 62 &  4.3\% \\
&P5: CE->FE->plausible  & 19 & 1.3\%  \\
&P6: FE->CE->plausible & 335 & 23.4\%  \\
&P7: FE->FE->plausible & 398 & \textbf{27.8\%} \\

\hline
Total & - & 1432 & 100\% \\

\hline
\end{tabular}


\label{table-patch-evolution}
\end{table}

\begin{mdframed}
Answer  to  RQ1:
\approach correctly fixes 74 bugs for 10 open-source projects from Defects4J 2.0, which outperforms our strong baselines from previous research with fault localization.
This state-of-the-art performance is made possible by the novel concept of iterative inference, providing the unique capability to iteratively improve partial patches towards final correct patches.
\end{mdframed}



\subsection{Answers to RQ2: Multi-location Bugs}

In RQ2, we focus on multi-location bugs, which is arguably one of the frontiers of program repair research.
\autoref{table-multi-location} gives the evaluation of \approach in repairing 35 multi-location bugs from Defects4J, with a quantitative comparison against VarFix \cite{varfix}, Hercules \cite{hercules}  \ICSERevision{and DEAR \cite{dear-icse22}}.
The first fourth columns show the bug information including their bug id, the number of classes that required edits, the number of edit locations in the ground truth patches, and the number of failing tests.
The repair results are shown in the last \ICSERevision{four} columns, where a  $\checkmark$ indicates the bug is successfully repaired and a  \colorbox{lightgray}{$\checkmark$} indicates the bugs only repaired by \approach and not by any technique from previous research.
In the last row, we summarize the repair results by showing the total number of repaired bugs. 

\textbf{Effectiveness} Over the \ICSERevision{35} evaluated multi-location bugs, \approach repairs \ICSERevision{17} of them with \ICSERevision{9} unique bugs that have never been repaired in the literature before.  This result outperforms Hercules by \ICSERevision{41.6\% (17 vs. 12), VarFix by  240\% (17 vs. 5) and DEAR by 750\% (17 vs. 2)}. We explain this major improvement of \approach as follows.
First, \ICSERevision{none of the three related  work } conducts any kind of patch improvement or iterations, \approach is the first approach to iteratively repair bugs with neural networks. 
This concept of iterative inference is essential to stack edit locations one after the other and synthesize a final multi-location patch. 
Second, \approach is the only approach that re-executes fault localization after a partial patch is found that reduces the number of failing tests. This key technical feature eliminates the suspicious locations that have been repaired, and enables the other buggy locations involved in the multi-location bug to be identified and better ranked, as more suspicious.


\begin{figure}
\centering
\includegraphics[width=0.48\textwidth]{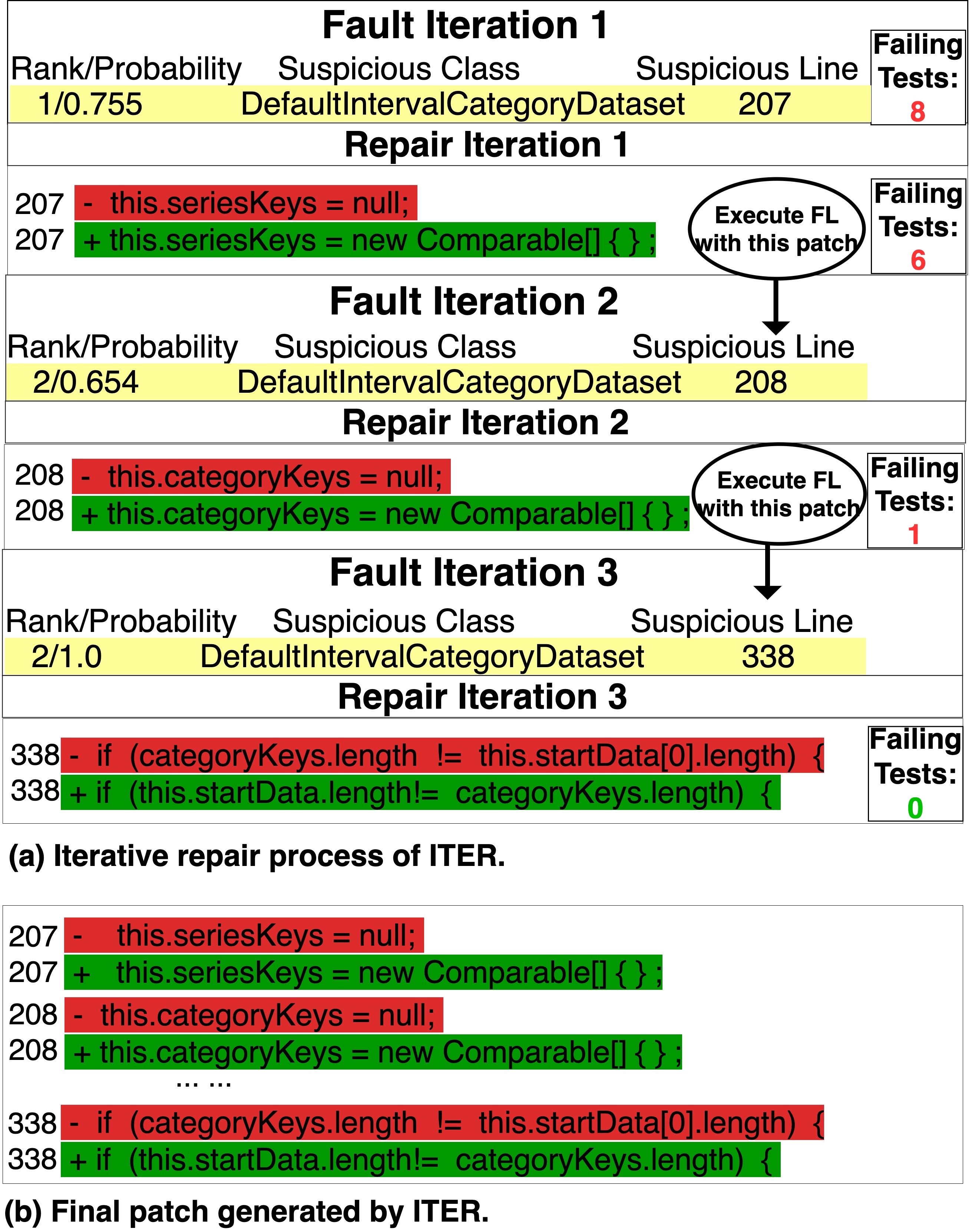} 
 \caption{A multi-location patch uniquely found by \approach for Chart-16.}
\label{fig:multi-bug-repair-example}
\end{figure}

\textbf{Case Study.}
\autoref{fig:multi-bug-repair-example} gives a unique patch for bug \texttt{Chart-16}, that is only found by \approach. Bug \texttt{Chart-16} is a multi-location bug because it requires three fixes.
\autoref{fig:multi-bug-repair-example} (a) shows the iterative process of \approach in generating this patch and (b) gives a final patch generated by \approach after several iterations.
\approach starts the repair iteration based on the suspicious \texttt{line 207} identified in suspicious class \texttt{DefaultIntervalCategoryDataset}, ranked in the 1st place with a suspicious probability score of \texttt{0.755}. 
After one repair iteration, \approach obtains a partial patch that reduces the original \texttt{8} failing tests to \texttt{6}  failing tests. 
According to the design of \approach (see line 23 in Algorithm \autoref{alg:inference}), \approach re-executes fault localization in the next repair iteration and identifies new suspicious locations. 
In the second repair iteration, \approach obtains another partial patch that reduces the previous \texttt{6} failing tests to \texttt{1} single failing test, which results in re-executing the fault localization again.
Finally, \approach generates a patch that passes all test cases. This enables \approach to succeed in generating the patch for \texttt{Chart-16}.

This bug cannot be repaired by VarFix due to the lack of appropriate suspicious locations as they only execute fault localization once. 
Hercules by its design cannot repair this bug as Hercules only searches for similar-looking code to apply the same fix in different locations, while here the patch is not composed of similar code.

\begin{table}[th!]
\renewcommand{\arraystretch}{0.9}
\footnotesize
\caption{Multi-location Bug Repair. \approach improves over the related work that specifically targets multi-location repair.}

\begin{tabular}[t!]
{p{0.165\linewidth}p{0.06\linewidth}p{0.06\linewidth}p{0.06\linewidth}cccp{0.07\linewidth}}

\hline
\multicolumn{4}{c}{Bugs} & \multirow{3}{*}{\textbf{VarFix}}  &   \multirow{3}{*}{\textbf{ Hercules}} & \multirow{3}{*}{\textbf{ \ICSERevision{DEAR}}}    & \multirow{3}{*}{\textbf{ITER}} \\
\cline{1-4} \\
ID &Classes & Edits  &  Failing Tests &   \\
\hline
Chart-14	& 2 & 4	 & 4 & &  \checkmark & & \\
\ICSERevision{Chart-16}	& \ICSERevision{1} & \ICSERevision{3}	&  \ICSERevision{8} & &  & & \colorbox{lightgray}{\ICSERevision{\checkmark}} \\

\ICSERevision{Cli-34}	& \ICSERevision{1} & \ICSERevision{2}	&  \ICSERevision{2} & &  & & \colorbox{lightgray}{\ICSERevision{\checkmark}} \\

Closure-4 &1 & 2 & 2	& & \checkmark	& & \checkmark  \\
Closure-78 &1 & 2 & 1	&	&  & &   \\
Closure-79 & 2 & 2 & 5  & &  & & \colorbox{lightgray}{\checkmark} \\
Closure-106 & 2 & 2 & 4	&  &  &&  \\
Closure-109 &1 &2 &2 &	   & \checkmark &&  \\
Closure-115	&1 & 2 & 7 &  	 &  &&  \colorbox{lightgray}{\checkmark} \\
Closure-131	& 1 & 2 & 2&	 	 &  && \colorbox{lightgray}{\checkmark} \\

\ICSERevision{Codec-1}	& \ICSERevision{3} & \ICSERevision{3}	&  \ICSERevision{5} & &  & & \colorbox{lightgray}{\ICSERevision{\checkmark}} \\

Lang-7	& 1 & 2 & 1 &	& & & \\
Lang-27	& 1& 2 &1	&  &  &  \\
Lang-34	& 1 & 2 & 27& 	 &  && \colorbox{lightgray}{\checkmark} \\
Lang-35	&1 & 2 & 1	&  &  & & \\
Lang-41	&1 & 4 & 2	&  &  & & \\
Lang-47	&1 & 2 & 2	&  & \checkmark& & \checkmark \\
Lang-60 & 1 & 2 & 1	&   & \checkmark && 	\\
Math-1	&2 & 2 & 2	&   &     & & \colorbox{lightgray}{\checkmark}   \\
Math-4	&2 &2& 2	&  & \checkmark&& \checkmark	\\
Math-22	&2 & 2 & 2&	  \checkmark &   & \checkmark  & \checkmark   \\
Math-24	&1 & 2 & 1&	 & \checkmark&& 	\\
Math-35	&1 & 2 & 4&	   \checkmark  & \checkmark&& \checkmark	\\
Math-43	&1 & 3 & 6&	&  \checkmark& &\checkmark	\\
Math-46	&	1 & 2 & 2	& \checkmark & \checkmark&& \checkmark	\\
Math-62	&	1 & 3 & 1	  & \checkmark& &	&\\
Math-65	&1 & 2 & 1	&& && \\
Math-71	&2 & 2 & 2&	&  && \\
Math-77	&2 & 2 & 2&	& && \colorbox{lightgray}{ \checkmark}	\\
Math-79	& 1&2 & 1&	 & & &	\\
Math-88	& 1&3 & 1&	 \checkmark& & &	\\
Math-90	&1 & 2 &1&	&  & 	&\\
Math-93	&1 & 4 & 1& &  & &	\\
Math-98	&2 & 2 & 2&	& \checkmark&  \checkmark  &\checkmark	\\
Math-99	&1 & 2 & 2&	 & & &	\\
\hline

 \multicolumn{4}{c}{  \multirow{2}{*}{ Total 35 bugs}  }    & \multirow{2}{*}{ 5 } & \multirow{2}{*}{ 12 } &  \ICSERevision{\multirow{2}{*}{ 2 } }    &  \multirow{2}{*}{ \colorbox{lightgray}{ \ICSERevision{17} }} \\

\\
\hline

\hline

\end{tabular}

\label{table-multi-location}
\end{table}

\begin{table*}
\renewcommand{\arraystretch}{1.1}
\footnotesize
\caption{
\ICSERevision{Detailed Analysis of Fault Localization.  The \checkmark and \ding{55} ~respectively indicates the true location is
selected or not selected for repair edits, while the '-' symbol denotes that the edit location is not applicable.
}
}

\begin{tabular}
{p{0.2\linewidth}|lll|lll|lll|c|ccccccccccc}

\hline

\multirow{2}{*}{\textbf{\ICSERevision{Bugs}}} &  \multicolumn{3}{c}{ \multirow{2}{*}{\textbf{\ICSERevision{DEAR FL}}}}  & \multicolumn{3}{c}{\multirow{2}{*}{\textbf{\ICSERevision{ITER w/o iterative FL}}}}   & \multicolumn{4}{c}{ \multirow{2}{*}{\textbf{\ICSERevision{ITER Iterative FL}}} }\\
&\\
& \multicolumn{1}{c}{ \ICSERevision{Edit1} }&\multicolumn{1}{c}{\ICSERevision{Edit2}} & \multicolumn{1}{c}{\ICSERevision{Edit3}}  & \multicolumn{1}{c}{ \ICSERevision{Edit1} }&\multicolumn{1}{c}{\ICSERevision{Edit2}} & \multicolumn{1}{c}{\ICSERevision{Edit3}}  &\multicolumn{1}{c}{ \ICSERevision{Edit1} }&\multicolumn{1}{c}{\ICSERevision{Edit2}} & \multicolumn{1}{c}{\ICSERevision{Edit3}}  & \ICSERevision{Uniquely Repaired}\\


\hline
\ICSERevision{Chart-16}  &  \ICSERevision{\ding{55} } & \ICSERevision{\ding{55} }   & \ICSERevision{\checkmark}      &  \ICSERevision{1st} &\ICSERevision{2nd} & \ICSERevision{19th}  & \ICSERevision{1st} & \ICSERevision{2nd} & \ICSERevision{2nd} (\textcolor{green}{$\uparrow$})  &   \\

\ICSERevision{Cli-34} &\ICSERevision{\ding{55} }  &\ICSERevision{\ding{55} }  &\ICSERevision{-} & \ICSERevision{42nd} & \ICSERevision{\ding{55} }    &\ICSERevision{-} &  \ICSERevision{42nd} &   \ICSERevision{3rd} (\textcolor{green}{$\uparrow$}) &\ICSERevision{-} &  \ICSERevision{\checkmark}  \\

\ICSERevision{Closure-4} & \ICSERevision{\ding{55} }  &\ICSERevision{\ding{55} }  &\ICSERevision{-} &  \ICSERevision{35th} & \ICSERevision{\ding{55} } &\ICSERevision{-} &  \ICSERevision{35th}  &  \ICSERevision{39th} (\textcolor{green}{$\uparrow$}) &\ICSERevision{-} & \ICSERevision{\checkmark}   \\

\ICSERevision{Closure-79} & \ICSERevision{\ding{55} }   & \ICSERevision{\ding{55} }   &\ICSERevision{-} &\ICSERevision{6th} &    \ICSERevision{\ding{55} }  &\ICSERevision{-}    & \ICSERevision{6th} &  \ICSERevision{4th} (\textcolor{green}{$\uparrow$}) & \ICSERevision{-}  & \ICSERevision{\checkmark} \\

\ICSERevision{Closure-115}   & \ICSERevision{\ding{55} }  &   \ICSERevision{\ding{55} }    &  \ICSERevision{-} &\ICSERevision{1st}  &\ICSERevision{10th}  &\ICSERevision{-} &\ICSERevision{1st} &\ICSERevision{5th}(\textcolor{green}{$\uparrow$})	&\ICSERevision{-}\\

\ICSERevision{Closure-131} & \ICSERevision{\ding{55} }  &   \ICSERevision{\ding{55} }    &  \ICSERevision{-} & \ICSERevision{16th} & \ICSERevision{\ding{55} }  &\ICSERevision{-} &\ICSERevision{16th} &\ICSERevision{4th}  (\textcolor{green}{$\uparrow$})   &\ICSERevision{-} &    \ICSERevision{\checkmark}    \\

\ICSERevision{Codec-1} &\ICSERevision{\ding{55} }  &\ICSERevision{\ding{55} }  &\ICSERevision{\ding{55} } &  \ICSERevision{1st}    &   \ICSERevision{3rd}
& \ICSERevision{\ding{55} } & \ICSERevision{1st} & \ICSERevision{1st} (\textcolor{green}{$\uparrow$}) &  \ICSERevision{1st} (\textcolor{green}{$\uparrow$})  &  \ICSERevision{\checkmark}   &     \\

\ICSERevision{Lang-34}	&\ICSERevision{\ding{55} }  &\ICSERevision{\ding{55} }  &\ICSERevision{-}  &  \ICSERevision{13th}  & \ICSERevision{15th} &\ICSERevision{-} & \ICSERevision{13th}  &\ICSERevision{28th}(\textcolor{red}{$\downarrow$})  &\ICSERevision{-}  \\

\ICSERevision{Lang-47}  &\ICSERevision{\ding{55} }  &\ICSERevision{\ding{55} }  &\ICSERevision{-}  &  \ICSERevision{5th}  &\ICSERevision{9th}  &\ICSERevision{-} &\ICSERevision{5th} &\ICSERevision{4th}(\textcolor{green}{$\uparrow$})  &  \\

\ICSERevision{Math-1}	&\ICSERevision{\ding{55} }  &\ICSERevision{\ding{55} }  &\ICSERevision{-} &  \ICSERevision{1st} &  \ICSERevision{8th} &\ICSERevision{-} &  \ICSERevision{1st} &   \ICSERevision{3rd} (\textcolor{green}{$\uparrow$}) &\ICSERevision{-} &   \\

\ICSERevision{Math-4}	&\ICSERevision{\ding{55} }  &\ICSERevision{\ding{55} }  &\ICSERevision{-} &  \ICSERevision{4th} &  \ICSERevision{\ding{55} }  &\ICSERevision{-} &  \ICSERevision{4th} & \ICSERevision{6th} (\textcolor{green}{$\uparrow$})  &\ICSERevision{-} & \ICSERevision{\checkmark}   \\

\ICSERevision{Math-22}	 &\ICSERevision{\checkmark}  &\ICSERevision{\checkmark}  &\ICSERevision{-}  &  \ICSERevision{1st} &  \ICSERevision{2nd} &\ICSERevision{-} &  \ICSERevision{1st} &  \ICSERevision{1st} (\textcolor{green}{$\uparrow$})&\ICSERevision{-} &  &  \\

\ICSERevision{Math-35} &\ICSERevision{\ding{55} }  &\ICSERevision{\ding{55} }  &\ICSERevision{-}  & \ICSERevision{3rd} & \ICSERevision{7th} &\ICSERevision{-} & \ICSERevision{3rd} & \ICSERevision{7th}(\tiny{$\rightarrow$}) &\ICSERevision{-} &  \\      

\ICSERevision{Math-43} &\ICSERevision{\ding{55} }  &\ICSERevision{\ding{55} }  &\ICSERevision{\ding{55} }	&\ICSERevision{35th} &\ICSERevision{36th}  &\ICSERevision{3\ICSERevision{7th}} &\ICSERevision{35th} & \ICSERevision{36th} (\tiny{$\rightarrow$}) &\ICSERevision{3\ICSERevision{7th}} (\tiny{$\rightarrow$}) &  \\

\ICSERevision{Math-46}	 &\ICSERevision{\ding{55} }  &\ICSERevision{\ding{55} }  &\ICSERevision{-} &  \ICSERevision{1st} &  \ICSERevision{\ding{55} } &\ICSERevision{-} &\ICSERevision{1st} & \ICSERevision{1st} (\textcolor{green}{$\uparrow$})  &- & \ICSERevision{\checkmark}  	\\

\ICSERevision{Math-77}	&\ICSERevision{\ding{55} }  &\ICSERevision{\ding{55} }  &\ICSERevision{-}  &	\ICSERevision{1st}  &  \ICSERevision{17th}  &\ICSERevision{-} & \ICSERevision{1st} & \ICSERevision{11th} (\textcolor{green}{$\uparrow$}) &\ICSERevision{-} &  \\

\ICSERevision{Math-98}  &\ICSERevision{\checkmark}  &\ICSERevision{\checkmark}  &\ICSERevision{-}	& \ICSERevision{4th}  & \ICSERevision{14th} &\ICSERevision{-} &\ICSERevision{4th} &\ICSERevision{4th}(\textcolor{green}{$\uparrow$}) &\ICSERevision{-}&  	\\

\hline

\ICSERevision{Number of Repaired Bugs} &\multicolumn{3}{c}{ \ICSERevision{2} }& \multicolumn{3}{c}{\ICSERevision{10} }& \multicolumn{3}{c}{\ICSERevision{17}} \\

\hline

\end{tabular}

\label{table-multi-location-ablation-study}
\end{table*}

\ICSERevision{Let us now discuss the performance of DEAR, which aims to find out multiple related suspicious locations and  repair them at once with  deep learning models.}
\ICSERevision{\autoref{table-multi-location-ablation-study}
gives a comprehensive analysis of fault localization  for the 17 repaired bugs. 
The analysis compares the fault localization results obtained from DEAR, \approach  without iterative fault localization (FL), and \approach  with iterative FL.
In the table, the  \checkmark and \ding{55} ~ 
respectively  indicates the true location is selected or not selected for repair edits, while the numbers indicate the ranked position for the corresponding edit location. 
The \textcolor{green}{$\uparrow$}, \textcolor{red}{$\downarrow$}, and $\rightarrow$ symbols signify whether the ranked position  improves, decreases, or remains unchanged for \approach with iterative FL. 
The \checkmark~ in the last column indicates the bugs uniquely repaired with iterative FL. 
In the final row, we summarize the number of repaired bugs based on the three considered analyses.  
\approach successfully includes all correct edit locations for 17 bugs, whereas DEAR only selects for two bugs (Math-22 and Math-98), and from this, we draw the following observations.
}

\ICSERevision{
Our experiments demonstrate that ITER, both without and with iterative fault localization, significantly outperforms DEAR thanks to the concept of partial patches.  DEAR heavily relies on identifying the correct combination of pair-wise edit locations and aims to fix them together in a simultaneous manner. In contrast, ITER does not rely on pair-wise locations and instead leverages the partial patches to make decisions to keep or discard an edit location. Consequently,  DEAR redundantly makes multiple repair attempts for a single edit location, depending on how many times it is combined with other different edit locations. On the other hand, ITER repairs each location independently.  If ITER succeeds in producing a partial patch for one edit location, the partial patch is retained and that location is not revisited again. This enables ITER  to mitigate the pair-wise exponential explosion.
}

\ICSERevision{In theory, the likelihood for DEAR to  identify correct pair-wise locations from a large combination of pairs is indeed low. For instance, when considering the top-50 suspicious edit locations suggested by SBFL, the probability of correctly identifying the pair is only $\frac{1}{\Comb{50}{2}}$ (0.08\%). 
This is confirmed in practice, resulting in the identification of only two successful pairs of edit locations. }


\ICSERevision{\approach with iterative fault localization demonstrates a substantial 
improvement by repairing 7 more bugs compared to \approach  without iterative FL.
This is because of the failure of simple SBFL to include the edit locations in the ranked list of considered suspicious locations. 
However, upon re-executing FL with partial patches, these previously neglected edit locations emerge in the top list.
Notably, over 17 bugs, iterative FL  enhances the ranking of edit locations for 16 bugs by re-ranking the edit locations in  earlier positions or emerge (\textcolor{green}{$\uparrow$}). 
This refinement in FL enables \approach  to faster locate the correct patches by providing more accurate edit locations.
}

\begin{table}[t!]
\renewcommand{\arraystretch}{0.98}
\footnotesize
\caption{Ablation Study of \approach w.r.t Training and Inference Iterations. The numbers are the plausible patches.}

\begin{tabular}
{p{0.24\linewidth}p{0.25\linewidth}p{0.09\linewidth}p{0.07\linewidth}p{0.07\linewidth}ccc}

\hline


& & \multicolumn{3}{c}{Inference Iteration} \\
\cline{3-5}
Training Iteration&  Training Samples &1 &2 & 3  \\ 

\hline

$\approach_{0}$ & \numprint{651787} & 68   & 93 & 104\\
$\approach_{1}$ &   \numprint{1192979} & 71  & 101 & 113 \\
$\approach_{2}$ &  \numprint{2071991}  & 72   &  104   & 117 \\
\hline
$\approach$ &  \numprint{3377864} &   79  &   105  & 119 \\

\hline

\end{tabular}

\label{ablation-study}
\end{table}

\textbf{Analysis of Unrepaired Bugs}
Over 35 evaluated bugs, \approach fails to repair 17 of them and we summarize the following  limitations. First, there are 12 bugs \approach that cannot be repaired due to the single failing test. \approach requires the signal of failing test cases to keep the partial patch and locate different locations. 
 Second, there are 6 bugs \approach that cannot be repaired due to the complexity of partial patches. \approach only conducts three iterations and fails to generate the correct fixes at all different locations. This suggests increasing the iteration number would be possible to repair more multi-location bugs.

\begin{mdframed}
Answer to RQ2:
\approach outperforms all multi-location repair approaches.
This demonstrates the power of iterative repair. Re-execution of fault localization is the key to correctly locating suspicious locations to be repaired in a meaningful sequence. 
\ICSERevision{Nine (9)}
multi-location bugs repaired by \approach were never reported as fixed in previous literature.
\end{mdframed}

\subsection{Answers to RQ3: Ablation Study}
In this RQ, we do an ablation study regarding the importance of the number of training iterations. We evaluate  \approach with a different number of training iterations. 
\autoref{ablation-study} gives the evaluation results on 476 bugs from 10 open-source projects in Defects4J 2.0.
It is the matrix where the columns represent the number of training iterations while the rows capture the number of inference iterations.
In addition, the second column gives the number of generated self-augment training samples.
Each cell indicates the number of bugs that have been repaired with plausible patches.

As expected, the training iterations do increase the number of generated training samples, approx. 2x for every training iteration.
Regarding performance, we make the following observations.
First, more training iterations with more training samples indeed improve the effectiveness of \approach (performance monotonously improves column-wise).
Second, looking at the first row (no training iteration, one single pass of training sample generation), we see that the inference iteration alone improves the effectiveness of \approach. This suggests all neural repair approaches would potentially benefit from iterative inference.
Third, the number of inference iterations contributes more than the number of training iterations.  For example, by using three inference iterations, \approach boosts the plausibility number from 79 to 119 (last row), which is a 25.2\% improvement. Lower gains are observed across columns, with for example a 14.4\%  improvement in the last column (104 to 119).

\begin{mdframed}
Answer to RQ3:
\approach introduces two novel concepts for program repair, iterative training and iterative inference. Our experiments show that both contribute to the final effectiveness of \approach. Yet, iterative inference makes a bigger contribution than iterative training. This suggests that most related work in program repair can achieve better performance by using this potent concept at inference time. This remark applies to the recent promising results on performing program repair with large language models \cite{xia-llm-icse23,alpha-repair,jiang2023knod,fan-automatedicse23}.
\end{mdframed}

\subsection{Answers to RQ4: Time Cost}

\begin{figure}[t!]
\includegraphics[width=0.42\textwidth, height=4.2cm]{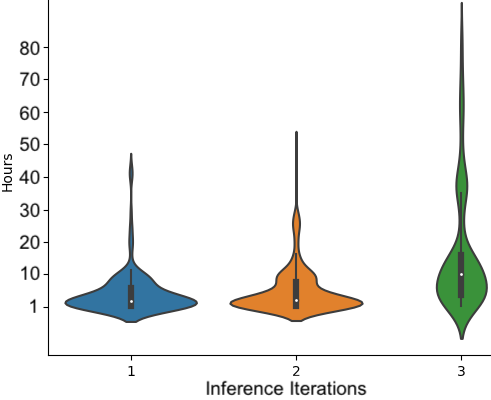} 

 \caption{Time cost of \approach in generating plausible patches.}

\label{fig:timecost}
\end{figure}

\autoref{fig:timecost} shows the time cost distribution of 1432 plausible patches discussed in RQ1.
For a single patch, the overall time cost stacks neural net usage, compiler and test execution and fault localization.
The x-axis indicates the number of inference iterations and the y-axis indicates the hours spent in generating the plausible patches. 

The majority of plausible patches are produced in the range of 1 to 10 hours.
Per the  concept of iterative inference, the time cost increase with the number of inference iterations, both in terms of median and outliers. 
The full quantitative analysis of the time cost is given \ICSERevision{in} \autoref{table-time-cost}, where the first three columns show the minimum, maximum and median time that \approach costs. The range is from \texttt{0.0053} hours to at most \texttt{84.5} hours. We can see half of the plausible patches are generated within \texttt{4.57} hours (median in the third column). In addition, we compute that 18.7\% and 33.2\% of plausible patches are generated within respectively one and two hours. This is significant per the study by Noller et al. \cite{nollertrust-icse22} showing that patches generated within one hour are appreciated by developers in practice.

We give those end-to-end results because they are the ones that matter to practitioners.
We note that these numbers cannot be directly compared with data from related papers because they tend to exclude the fault localization time and patch validation time.
One reason for that is many experiments assume perfect fault localization (e.g., Xia et al. \cite{xia-llm-icse23} reported that the LLM takes 2.5 hours to generate 600 candidate patches for Defects4J with perfect fault localization).  Therefore, we argue that our measurements are more meaningful both from a scientific and industrial perspective.

\begin{mdframed}
Answer to RQ4:
Half of the plausible patches from \approach are generated within 4 hours and a half.
18.72\% of plausible patches can be generated within one hour.
\end{mdframed}

\section{Threats to Validity}

An internal threat relates to 1) our fault localization tool GZoltar may cause failures for some evaluated bugs due to dependency and version issues, which prevents \approach from generating more plausible patches.
2) manual patch correctness assessment. To mitigate these threats, we make our tool and generated patches publicly available.

Per the standards of the field, our approach has been tested in one language (Java) and the evaluation is carried out on well-established benchmarks. 
A threat to external validity relates to whether the novel concepts of \approach, iterative training, and iterative inference generalize to arbitrary programming languages. We do believe that our approach can be applied to other programming languages and datasets and that performance would follow accordingly.

\section{Related work}






\subsection{Program Repair}


\ICSERevision{
Over the past decade, extensive research efforts have yielded a substantial body of work on automatic program repair, which can be broadly categorized into three categories \cite{goues2019automated}: search-based repairs, semantic-based repairs and learning-based repairs.
Search-based repair approaches, such as GenProg \cite{LeGoues2012GenProg}, Relifix \cite{relifix-shinwei-icse15}, ssFix \cite{ssFix}, SoFix \cite{sofix}, and others \cite{jaid,Yuan2017ARJAAR,astor,ali-issta19-bytecode,sharpFix} typically define a search space with  different kinds of edit patterns, such as inserting null checks \cite{NPEFix}, copying statement \cite{capgen-ICSE18}, mutating operators \cite{OperatorMutation}, or modifying expressions \cite{s3,tbar}. Search-based repair approaches first generate candidate patches and then they use a search algorithm to locate plausible patches from patch search space.
Contrary to search-based repair generation approaches,  semantic-based program repair techniques, such as Nopol \cite{nopol}, SemFix \cite{semfix} and others \cite{acs,JFix, concolic-repair-PLDI21,patch-transplantation-TOSEM21,CrashProgramRepair-ISSTA19} first construct constraints     
that should be satisfied to fix a bug, and then they use program synthesis to synthesize patches that satisfy the repair constraints.
}


\ICSERevision{
%
Learning-based repair uses neural networks, and has been mostly with supervised learning on commits. 
Supervised learning-based repair approaches, such as Codit \cite{codit-tse20}, Modit \cite{modit-ase21}, CURE \cite{CURE-icse21}, DLFix \cite{DLFix} and Tufano's work \cite{Tufano-tse19,Tufano-ICSE19} learn to generate fixes from past commits submitted by developers: given a buggy code and surrounding context, the neural network is optimized to translate the buggy code to the correct code.
In essence, supervised learning-based repair learns code transformations from a vast number of past commits, e.g., open-source projects in GitHub \cite{Tufano-ICSE19,SEQUENCER,CoCoNuT,CURE-icse21}.
Contrary to the above supervised learning approaches,  another line of work \cite{selfAPR,allamanis2021self-buglab,Yasunaga20DrRepair,BIFI-ICML21} has proposed to use self-supervised learning to completely get rid of commit collection in the learning process.
}

\ICSERevision{There are two key differences between ITER and the aforementioned works. First, none of them conducts iterative refinement with partial patches, by chaining repairs of compilation errors and test failure errors. Second, no existing work re-executes fault localization with  partial patch and demonstrates major empirical results on multi-location bugs.
}

\subsection{Patch Evolution}
\ICSERevision{
Prior work considered patch evolution into the repair process,  which is related to the idea of iterative repair presented in this paper. For example, Gupta  et al. \cite{deepfix} propose DeepFix to iteratively repair compilation errors.  ITER is different from DeepFix as follows.  First, DeepFix's iterative process only proceeds  when all previous repairs are accepted by the compiler (i.e., compilable). If any patch is rejected and deemed  non-compilable, DeepFix's iterative process is terminated. In contrast,  ITER conducts novel patch refinement through multiple attempts  to even improve low-quality patches.  Second, DeepFix does not incorporate a fault localization process since it focuses on resolving compilation errors, while one of the key components in ITER is the iterative fault localization process with partial patches. 
}

\ICSERevision{
Arcuri and Yao \cite{co-evolutionary-Arcuri-Yao-2008} present a co-evolutionary approach to automatically fix software bugs.  Their approach utilizes genetic programming to evolve programs that pass  the existing set of unit tests, while also employing test generation techniques yield new tests for finding bugs in the evolutionary programs.
ITER shares a similar concept of patch evolution.
However, there are some notable differences. 
Arcuri and Yao's work  assumes a known formal specification, allowing for the generation of potentially infinite new test cases.
On the other hand, 
ITER considers a fixed number of test cases written by developers from the real-world, without assuming any high-level specification. 
In addition, ITER incorporates an iterative fault  localization process which is not considered in their work.
}

It is worth mentioning that GenProg \cite{genpro2009,LeGoues2012GenProg}, over a decade ago,  considered an iterative process of stacking edits by a genetic programming algorithm. 
Yet, we revisit this idea with major novelty.
First, GenProg does not consider partial patches with compilation errors which are simply discarded. \approach transforms uncompilable patches into plausible and correct ones.
Second, GenProg does not re-execute fault localization for partial patches. \approach demonstrates that this is essential for multi-location bugs.
Third, GenProg is based on AST transformation, while \approach  builds up on deep learning for repair that did not exist at GenProg time and has been shown  
powerful in recent years.

\begin{table}[t!]
\renewcommand{\arraystretch}{1.18}
\footnotesize
\caption{Time cost analysis of \approach in generating 1432 plausible patches}

\begin{tabular}{cccccccc}

\hline

 \multicolumn{3}{c}{Time}  & & \multicolumn{2}{c}{Proportion}\\
\cline{1-3}\cline{5-6}
min  & max  & median & &  < 1 Hour & < 2 Hours \\
\hline
0.0054 H & 84.50 H & 4.57 H & & 18.72 \%  & 33.20\% \\
\hline

\end{tabular}

\label{table-time-cost}
\end{table}



\subsection{Multi-location Program Repair}
In practice, most APR approaches do not aim to multi-location repair, and there are only a few works that target multi-location repair based on 
\ICSERevision{ symbolic variables and }
heuristics for patch location combination.
\ICSERevision{
Semantic program repair techniques, such as DirectFix \cite{directfix} and Angelix \cite{Angelixicse16}, facilitate multiple-location fixes by substituting multiple suspicious expressions with multiple symbolic variables.
The synthesis of a multi-location patch involves replacing multiple symbolic variables with concrete values through the process of symbolic execution and constraint solving.
}
Wong et al. \cite{varfix}'s VarFix use variational execution to  merge all possible edits into a meta-program, where all edits are guarded by if-conditions that control whether to include the edit at runtime. 
Saha et al.'s Hercules  \cite{hercules} repair multi-location bugs  by searching for similar code in commit history to find so-called evolutionary siblings. In essence,  Hercules applies a substantially similar transformation at a number of different locations.
\ICSERevision{
Li et al. \cite{dear-icse22}'s DEAR is a deep learning-based approach, which selects and combines multiple buggy locations provided by SBFL. DEAR's objective is to generate patches for all identified locations simultaneously to repair multi-location bugs.}
\approach is fundamentally different.
First, \ICSERevision{none of approaches mentioned above} considers an iterative repair process to refine the generated patches and they do not re-execute fault localization to obtain updated suspicious locations. This is what we do, and we provide strong empirical evidence that this works.
Second, \approach is not limited to repairing multiple locations with similar edits as Saha et al. \cite{hercules}. Our experimental replication package contains multi-location patches where the edits are completely different one from another, yet combined in a meaningful multi-location patch.
\ICSERevision{
Third,  unlike Hercules and DEAR, ITER does not necessitate the combination of multiple buggy locations. Instead, ITER utilizes partial patches to determine whether to retain or discard each individual buggy location.
}


\section{Conclusion}
We have presented a novel neural program repair paradigm called iterative repair.
Our approach, called \approach, integrates fault localization, patch generation, and patch validation in an original iterative loop. 
This results in two breakthroughs:
First, \approach is able to improve partial patches that have been generated in prior repair attempts in order to converge to a final correct patch. 
Second, \approach is able to push the state-of-the-art on repairing multi-location bugs with 
\ICSERevision{9}
unique multi-location bugs that were never repaired before.
The original concepts \approach has wide applicability, incl. on maximizing repair performance with large language models.

Our future work focuses on improving the performance of \approach. To this end, we aim to investigate patch minimization to filter out less useful patches and not keep them for the next iteration. Having a good patch minimization technique will greatly improve the performance of \approach and benefit finding plausible patches at an early time.

\begin{acks}
We thank the anonymous reviewers for the insightful feedback.
This work was partially supported by The Wallenberg Foundation and WASP Postdoctoral Scholarship Program - KAW 2022.0368.
We thank Xuechun Xu from KTH Royal Institute of Technology for the support of computing resources.
Some experiments were performed on resources provided by the Swedish National Infrastructure for Computing.
\end{acks}

\newpage

\bibliographystyle{ACM-Reference-Format}
\bibliography{reference}

\end{document}